\documentclass{sig-alternate} 
\usepackage{times}
\usepackage{graphicx}
\usepackage{amssymb}
\usepackage{url,hyperref}
\usepackage{authblk}

\makeatletter
\def\@copyrightspace{\relax}
\makeatother

\author[1]{Jeremy Cohen}
\author[2]{Chris Cantwell}
\author[3]{Neil Chue Hong}
\author[4]{David Moxey}
\author[5]{\authorcr Malcolm Illingworth}
\author[5]{Andrew Turner}
\author[1]{John Darlington}
\author[4]{Spencer Sherwin}

\affil[1]{Department of Computing, Imperial College London, South Kensington Campus, London SW7 2AZ, UK}
\affil[2]{National Heart \& Lung Institute, Imperial College London, South Kensington Campus, London SW7 2AZ, UK}
\affil[3]{Software Sustainability Institute, University of Edinburgh, JCMB, Mayfield Road, Edinburgh EH9 3JZ, UK}
\affil[4]{Department of Aeronautics, Imperial College London, South Kensington Campus, London SW7 2AZ, UK}
\affil[5]{EPCC, University of Edinburgh, James Clerk Maxwell Building, Mayfield Road, Edinburgh EH9 3JZ, UK}

\begin{document}

\title{Simplifying the Development, Use and Sustainability of HPC Software}

\maketitle
\thispagestyle{empty}

\begin{abstract}
   Developing software to undertake complex, compute-intensive scientific processes requires a challenging combination of both specialist domain knowledge and software development skills to convert this knowledge into efficient code. As computational platforms become increasingly heterogeneous and newer types of platform such as Infrastructure-as-a-Service (IaaS) cloud computing become more widely accepted for HPC computations, scientists require more support from computer scientists and resource providers to develop efficient code and make optimal use of the resources available to them. As part of the libhpc stage 1 and 2 projects we are developing a framework to provide a richer means of job specification and efficient execution of complex scientific software on heterogeneous infrastructure. The use of such frameworks has implications for the sustainability of scientific software. In this paper we set out our developing understanding of these challenges based on work carried out in the libhpc project.
\end{abstract}

%\begin{figure}[!b]
%  \begin{center}
    %\includegraphics[width=3.5in]{figure.pdf}
%  \end{center}

%  \caption{\small Figure caption. To get a figure to span two
%      columns, use the environment figure* rather than figure.}
%  \label{fig-label}
%\end{figure}

\section{Introduction}
\label{section:introduction}

Building scientific software for use on HPC platforms can be a complex process bringing together the specialist domain knowledge of the scientists who are likely to be the end-users of the resulting software, method developers, computer scientists and resource providers. As platforms get more advanced and are increasingly heterogeneous, more collaboration is required between these groups in order to develop efficient software. In addition to small groups of local servers or larger cluster facilities, scientists may wish to make use of Infrastructure-as-a-Service (IaaS) cloud computing platforms to gain access to more cores than they have available locally, or to different types of hardware. 

As part of the libhpc stage 1 and 2 projects we are developing a framework to provide a richer means of job specification and efficient execution of complex scientific software on heterogeneous infrastructure. This work is motivated by the desire to make it easier for end-users to both describe the tasks they want to undertake and to make use of a wider range of computational infrastructure, in a more streamlined manner. Our work on the libhpc project serves as a basis for our views on managing the use of complex infrastructure and creating sustainable software, which are set out in this paper. One of the key focuses of our work is to release the tight coupling that often exists between the different entities in the software development process and replace these with a set of less interdependent processes. We aim to achieve this through the specification of clearly defined interfaces and supporting metadata structures against which code can be developed without the detailed interactions that were previously required with other groups. Metadata is used to capture information about software and hardware. In the case of software, metadata records capabilities, features and hardware requirements while in the case of hardware, it stores platform specifications and capabilities.

In this paper we set out our developing approach based on work carried out in the libhpc project and build on previous material presented in~\cite{ref:cohen12-escience}. Section~\ref{section:libhpc} provides an overview of the current libhpc model while details of related systems and approaches are covered in Section~\ref{section:related-work}. Section~\ref{section:scientific-software} provides more details of how we see new methods for developing complex distributed HPC codes contributing to the long-term sustainability and flexibility of software.

%An automated mapper that can select suitable software implementations and compute platforms

\section{The libhpc framework}
\label{section:libhpc}

%This section provides an overview of the libhpc framework that is being developed as part of the libhpc proejct~\cite{ref:libhpc}. In section~\ref{section"scientific-software} we set out our thoughts on how libhpc and related systems can aid the effective development of scientific software in an environment of heterogeneous resources. An overview of the libhpc framework and how it has developed from the earlier presentation in~\cite{cohen12-escience}.

The members of the libhpc project~\cite{ref:libhpc} are designing and building a framework to support the development and use of scientific HPC codes on heterogeneous infrastructure. The libhpc model brings together the concept of software components and co-ordination forms: higher-order functional constructs that operate on components and define patterns of control. The software components provide the means of carrying out computation while co-ordination forms, originally developed in work by Darlington et al.~\cite{ref:darlington95}, offer a powerful means of specifying complex patterns of control between software components enabling the building up of applications. 

Our software component model defines abstract components, which contain only metadata describing high-level capabilities of the component, and concrete components that augment the metadata with software code that carries out computation. This approach is taken to allow the automated selection of the most suitable concrete component(s), given a specification consisting of abstract components and hardware selection, enabling a user to develop a single application description but have it run efficiently on a range of different computational platforms.

Libhpc also defines a deployment layer that acts an abstraction layer for various different types of computational infrastructure; for example, PBS or Grid Engine-based clusters, IaaS clouds or standalone local resources. Metadata about hardware is recorded in our metadata store to enable methods to be matched to suitable platforms. Ultimately, this will enable the core of the libhpc framework, the \textit{mapper}, to select the most suitable software components to undertake a user-defined task and then to deploy these components to the most appropriate hardware platform to support a user's requirements, ensuring that optimal component implementations are used given an initial, high-level, user application specification.

While some applications may be able to take advantage of the full libhpc framework from application specification through to mapping, deployment and execution management, there are many circumstances where use of the full libhpc stack may not be appropriate and only some elements of the framework can be used. For example, many complex HPC codes are tightly coupled and it will not be practical to break them down into a series of components. In such cases, the software component will be coarse-grained -- an application will be a single component. In this case, it is still possible to provide several builds of the application targeted to different platforms and allow libhpc's mapper and deployment layer to dynamically select a suitable hardware platform and corresponding software implementation at runtime.

\section{Related work}
\label{section:related-work}

%Use this section to provide detail of related work and, in particular, to highlight some of the features of competing frameworks or services. What are these missing? What additional challenges do they present. What do we feel are the aspects that make libhpc different and what are the key points that we're aiming solve in the development of our framework?

Extensive research has been carried out over many years motivated by the desire to make software easier to develop and use, particularly as new computing hardware and infrastructure patterns emerge. There are research programmes that have focused on linking application scientists with computer scientists, funding a range of projects to assist with optimising specific codes, supporting frameworks or the underlying infrastructure on which these codes are run. Software efforts, such as in the area of workflows, aim to simplify the use of multiple codes or tools that may previously have been scripted locally and to enable the use of distributed services in place of local resources. Efforts have also been made to develop compile-time optimisation tools that can take user code, perhaps with some higher-level annotations or constructs, and produce optimal code targeted at a specific platform or distributed environment.

The UK e-Science Programme~\cite{ref:ukescience}, which ran for many years, developed a range of tools and services to support the use of Grid computing infrastructure and funded a range of interdisciplinary, collaborative projects focusing on making use of this Grid computing infrastructure easier for scientists across a range of domains. Workflow environments such as Taverna~\cite{ref:WoHaFe13} provide a means of executing workflows consisting of multiple components that may be available locally or as remote Web Services. In addition to generic workflow systems, many systems have been developed to assist users in specific domains. Bioinformatics is an example of this where systems such as Galaxy~\cite{ref:galaxy05} or VisTrails~\cite{ref:Callahan06} provide domain-specific features to improve the user experience.

With the emergence of cloud computing, including Infrastructure as a Service (IaaS) public cloud platforms such as Amazon EC2~\cite{ref:ec2} or RackSpace~\cite{ref:rackspace} and private cloud frameworks such as OpenStack~\cite{ref:openstack}, access to large-scale, remote, distributed infrastructure has become much easier. Other types of architectures such as GPUs and FPGAs provide manifold opportunities for improving code performance but at the cost of the complexity of porting or building new code. Heterogeneous hardware can also require learning different development approaches so frameworks such as OpenCL\texttrademark~\cite{ref:opencl}, which provides a C-based language for developing cross-platform code, and OpenACC~\cite{ref:openacc}, which uses compiler directives to specify code that should be executed on alternative hardware, have emerged to provide a common approach to developing code that can be executed on different platforms. Similarly, OP2~\cite{ref:op2} is a framework for running applications on clusters of GPUs or multi-core CPUs. While these language extensions and frameworks provide methods for low-level optimisation, they require skilled developers with an understanding of the underlying hardware in order to be used effectively. Alternatives that offer the potential of supporting a much wider range of developers are compile-time auto-tuning and runtime optimisation. Auto-tuning can be applied in the context of libraries that generate optimised code at build time to undertake their specific functionality, for example, the Optimised Sparse Kernel Interface (OSKI) Library~\cite{ref:oski} that optimises code for sparse matrix operations. Auto-tuning compilers such as Milepost GCC~\cite{ref:milepostgcc} offer a more general option using advanced functionality to produce compiled code that is optimised for the platform that they are building on.

\section{Improving scientific software}
\label{section:scientific-software}

\subsection{Overview and Challenges}

What constitutes improvement in computer software? For some individuals it is likely to be better performance, for others it may be ease of use while others may be interested in additional features, greater extensibility or options for customisation. For scientific software, improved accuracy, more realistic models or more advanced algorithms may be of importance. What constitutes improvement is also likely to differ depending on the individual in question. We consider five different profiles for individuals or groups involved in scientific computing:

\begin{itemize}
\item End user: The end user is likely to be a scientist with domain expertise but may not be an expert in software development, numerical methods or the use of computing infrastructure.
\item Method developer: Method developers have domain expertise and software development experience and specialise in formulating scientific algorithms.
\item Application developer: Application developers have extensive software development experience and knowledge of numerical methods, and an understanding of different models of computing infrastructure and how to target application code to this infrastructure effectively.
\item Framework developer: These developers are computer scientists who have a technical understanding of frameworks used to decouple an application from the infrastructure on which it is run. They understand how to bring together the outputs of the various groups and users in the system and to optimise these to support different models of infrastructure.
\item Infrastructure provider: Infrastructure providers provision and operate large-scale infrastructure. They understand the requirements of HPC code and can support the optimal targeting of code to their platform(s).
\end{itemize}

The way that code is built is key in ensuring long-term sustainability and ongoing improvement of software but the development of scientific software presents challenges for many reasons. First and foremost is the complexity of many scientific methods and algorithms and the need to map scientists' and method developers' understanding of these into efficient code. In some cases the difficulty of doing this results in code that does not operate efficiently or that is not portable. In other situations, a method developer may produce high-quality code but the detailed understanding of the science that is mapped into this code is almost impossible to re-create at a later date making it difficult to maintain and extend. The range of different computing platforms available requires decisions when designing an application, particularly where parallel code is being developed, and this may affect the ease of using the application on other types of platform in the future. We strongly believe that by augmenting code with metadata that provides details of how and why code is designed and built in a particular way, it is possible to capture some of the knowledge and understanding of developers, and to attach this to the code to provide long term benefits. Of course, good software development practice dictates the writing of documentation and comments within the code are one form of metadata. However, we believe that by adding some form of structure to this metadata, it will be possible to record more detailed information about code that can subsequently be read by automated tools in order to optimise the way that code is deployed and run.

%Why is building well-architected, efficient software of such great importance?
As both experiments and computing infrastructure increase in size, the amount of work undertaken to build and run code also increases. The costs associated with software development and management of infrastructure can be very large and if code is not built well, vast amounts of money can be wasted and results can be inaccurate or even useless. If code is well built, all the user groups identified above stand to benefit in the long term, and end users can more effectively utilise the resources available to them.
%
%A general overview of the aims for improving scientific software. Why is this important? Who benefits? What are the challenges?
%
To support both the development experience and the user experience, where complex distributed software is concerned, we consider that decoupling the interactions between the entities involved in software development can deliver more robust code and a faster pace of development.
%This is discussed in the next section.

\subsection{Decoupling the development process}

In the previous section, we introduced five groups involved in building and using software. When building distributed applications targeting a range of HPC platforms, it is generally necessary for there to be extensive interaction between these groups to ensure the software meets user requirements, has correct algorithmic implementations and is able to run efficiently on a particular computational infrastructure. This arrangement inherently hampers sustainability of the software since modifying any particular aspect would require discussion between all parties. For example, changes or additions to an algorithm within a scientific code could require additional coding by the method developer, optimisation by the application computer scientist for the relevant platform via consultation with the infrastructure provider, and notification to the user about the changes to the software interface. Similarly, for a new end-user to run the software on their local platform requires the application developer and possibly also a framework developer to consult with the infrastructure provider and method developer to optimise the code for the user's chosen platform. It is likely to be the case that some groups will not expect to be in direct communication; for example, it should not be necessary for method developers to communicate with framework developers building auto-tuning compilers.

The risk to sustainability comes from the interdependency of the different user groups. Loss of one of these stakeholders can prevent software being adapted to meet current and future needs, leading to the software becoming unmaintainable. Decoupling should reduce the risk of software becoming dormant, and the use of metadata is a means to formalise the decoupling. For example, the method developer might attach appropriate metadata to components of the algorithm (e.g. solve linear system for N vectors), allowing an application developer to provide an optimised implementation (direct solve, factorise and solve, iterative solve) without further consulting the method developer. If the infrastructure provider then attaches sufficient metadata describing the capabilities of the hardware, the application developer could essentially develop the optimised application code in isolation. An example of this kind of approach can be seen in the various implementations of OpenMP.

However, decoupling presents its own challenges; the main problem being how to define what metadata should be attached to the various algorithms, components and hardware platforms in an application independent manner. Metadata is essentially a more structured, enforced form of documentation, which historically has been used to the same effect, but the diversity of software may limit the efficacy of this approach. Customisation of metadata structure on a per-application basis may be required, leading to increased and potentially unmanageable complexity. Furthermore, migrating existing software to use this approach could be challenging and potentially require redesign of the software. However, new software could be developed with decoupling in mind, ensuring it is written in a component-oriented fashion and augmented with suitable metadata. At present, members of the libhpc project team are undertaking an exercise to identify the significant properties of various types of scientific code by surveying user groups. The results of this process will provide important input to the design of a cross-domain metadata schema designed to capture key software properties based on user requirements.

\subsection{Software development communities}

While frameworks such as libhpc can provide ways to more easily describe complex computational jobs and target a range of infrastructure, the code providing the scientific methods, and the code of the framework itself, need to be maintained. One of the most practical ways to build a critical mass of interest and support for a code and, hence, the potential for long-term sustainability, is to encourage the establishment of \emph{communities} around particular codes or projects representing groups of codes. The members of the community contribute to the development and maintenance of the software in a distributed, yet coordinated, fashion. Such an approach distributes knowledge regarding all aspects of the code, spanning the methods, optimisation and deployment, across the community such that the loss of any one member is far less likely to hamper the sustainability and maintainability of the code.

Communities can work well where a large number of people have a vested interest in a particular tool, and there are many such examples amongst the projects hosted on systems such as SourceForge and GitHub. In the case of large and high-profile projects, communities can be very powerful, often taking on extensive project management and development tasks and providing a means for discussion spanning all stakeholders. In contrast, community building around small-scale scientific projects in a narrow application domain can be challenging due to the comparatively small size of the total user community, meaning it is harder to attain the critical mass of interest to seed the development of a supporting community.

% Can alternative approaches to developing distributed software for heterogeneous resources result in the building of communities? Are there are enough users with similar requirements in common? How might the development of such communities be beneficial? What are the similarities across domains that can be built upon?

\subsection{Lessons from libhpc}

The ultimate aim of the work being carried out in the libhpc project is to improve the long-term sustainability and usability of scientific software. To make it easier for end-users with limited computing knowledge to run this software and to make use of different types of computing platforms will require changes across all aspects of scientific software development. For the majority of scientific codes, we believe the decoupling approach offers a viable route to sustainability. This is the strategy we have taken in libhpc by describing software as interchangeable components, augmented with metadata, and using coordination forms as the mechanism to compose components to form applications.

We have looked at a range of software tools and explored their suitability to be adapted into a decoupled model. Our experience has shown that trying to express existing code as fine-grained components is challenging. Large C or C++ codes, for example, are frequently tightly coupled internally and the work necessary to rebuild these codes into a series of loosely coupled components far outweighs the potential benefits of libhpc. Nonetheless, these applications can instead be represented as a single coarse-grained component representing the complete application. While this removes some possible benefits by enabling only the complete application code to be replaced with an alternative build or set of parameters to support a different hardware platform, rather than selecting suitable alternatives for a range of different fine-grained components, it still offers benefits in platform selection and application life-cycle management. We have tested this approach in libhpc.

One important benefit from the work on libhpc so far has been the extensive interaction between computer scientists, method developers and end-users. This knowledge sharing has served to highlight how different terminologies and development approaches can present challenges in large development teams but it has also enabled us to identify key interfaces between our groups which, in turn, has allowed us to identify the best use of effort within the groups to build up the framework. We feel that while libhpc seeks to reduce the necessary technical interactions between development groups, it allows the focus to be put back onto the scientific methods themselves rather than on how to make them work on a specific type of platform. We have observed this in the tests we have done within the project using our target scientific codes.

%We have developed Nekkloud, a web interface for deploying the Nektar++ finite element solver to clouds and clusters. Deployment plugins provide an abstraction for access to the various infrastructure resources, thereby decoupling the end-user from the infrastructure provider. The end-user is also decoupled from the method developer and computer scientist who prepare the optimised build of the code for different platforms.

%% % Need to look at higher level view of code rather than fined grained
%% % Diversity of software

%In this section we will provide some more general discussion on our view of the benefits of improving the way scientific software is built, who can use it and how it is maintained. What have we learned from our libhpc work so far? How can software component and metadata repositories help with the long term usability and sustainability of software? How might the development of a community around frameworks like libhpc help provide wider benefits to both users and providers of advanced computational infrastructure?

\subsection{Sustainability of advanced frameworks}

We have presented material about how the libhpc framework can help to support the sustainability of scientific applications by making them easier to run and maintain. There is, however, still the question of how such advanced frameworks are, themselves, maintained and sustained. Unlike conventional software, frameworks need the input of each of the stakeholders so decoupling the framework development and maintenance is counterproductive. There is a question around whether end users should be directly involved in the framework development or whether they should be users only. On one hand having end-users and method developers (or a set of representative users) involved will encourage them to buy-in to the framework, which itself is important for long-term sustainability and maintenance. On the other hand, such complex interactions may be difficult to manage if it is attempted to satisfy the demands of a broad and divergent range of end-users and method developers, which in turn will make sustainability less likely. Ultimately we consider that for a framework to be successful in the long-term, it is necessary for all the entities involved in its development or use to  clearly see the value of the framework and understand its benefits. We believe that the emergence of a large development-focused community around a framework presents the best opportunity to keep code maintained and to keep the system evolving and providing users with new and improved features.

% We can show that advanced frameworks are necessary to make the software development process more effective when working with complex modern infrastructure but what is the sustainability of these frameworks themselves? Are we making things more complex by having to have a team that maintains and sustains the framework in addition to learning the nuances of new architectures and how to target code to them?

% Of the 4 (5?) user types - who should be involved in sustainability of these frameworks. Two schools of thought - is user involved in framework - need end user to buy into the framework and understand the benefits of sustainability - maybe have a set of representative end users who assist with this. Whether end users/method developers are/are not involved in the framework. End-users will, at least, be running software that is using the framework. Interaction tough to manage - if the aims of users/method developers are spread too far we'll get divergent end-user / method developer aims which make sustainability harder.

\section{Conclusions and further work}
\label{section:conclusions}

In this paper we have presented an approach for improving the usability and sustainability of scientific software for a range of different stakeholders based on our experiences in the libhpc project and related work. The focus has been on a proposed decoupling of the dependencies that exist between groups in the traditional approach to developing large-scale scientific applications. By capturing as much information as possible about a user's requirements, the capabilities of scientific codes and hardware platforms and storing this in metadata repositories, advanced middleware can be provided to compose components containing code, identify suitable hardware platforms for running this code and then handle the process of deploying and running the code. This provides end users with much more flexibility in the types of hardware platforms that are accessible to them and the overall experience that they have in developing code themselves, or working with method developers to assist with this, and running this code to obtain scientific results.

As we continue our work in the libhpc project we are implementing more of the framework and developing demonstrators to show how many of the approaches discussed here can be realised. It is hoped that this work will serve to support users in a range of domains who are part of the project and to provide us with valuable feedback to assist in optimising our approaches to improving scientific software for those who build and use it.

\bibliographystyle{abbrv}
\bibliography{libhpc-WSSSPE13}

\begin{thebibliography}{10}

\bibitem{ref:libhpc}
{libhpc: Intelligent Component-based Development of HPC Applications}.
\newblock http://www.imperial.ac.uk/lesc/projects/libhpc.
\newblock accessed 1$^{st}$ September 2013.

\bibitem{ref:openacc}
{OpenACC Home}.
\newblock http://www.openacc-standard.org/, 2013.
\newblock accessed 1$^{st}$ September 2013.

\bibitem{ref:openstack}
{OpenStack Open Source Cloud Computing Software}.
\newblock http://www.openstack.org/, 2013.
\newblock accessed 1$^{st}$ September 2013.

\bibitem{ref:ec2}
{Amazon Web Services, Inc.}
\newblock {Amazon Elastic Compute Cloud (Amazon EC2)}.
\newblock http://aws.amazon.com/ec2, 2013.
\newblock accessed 1$^{st}$ September 2013.

\bibitem{ref:galaxy05}
{B. Giardine et al.}
\newblock Galaxy: a platform for interactive large-scale genome analysis.
\newblock {\em {Genome Research}}, 15(10):1451--1455, Oct. 2005.

\bibitem{ref:Callahan06}
S.~P. Callahan, J.~Freire, E.~Santos, C.~E. Scheidegger, C.~T. Silva, and H.~T.
  Vo.
\newblock {VisTrails}: visualization meets data management.
\newblock In {\em {Proceedings of the 2006 ACM SIGMOD International Conference
  on Management of Data}}, SIGMOD '06, pages 745--747, New York, NY, USA, 2006.
  ACM.

\bibitem{ref:cohen12-escience}
J.~Cohen, J.~Darlington, B.~Fuchs, D.~Moxey, C.~Cantwell, P.~Burovskiy,
  S.~Sherwin, and N.~C. Hong.
\newblock libhpc: Software sustainability and reuse through metadata
  preservation.
\newblock In {\em First Workshop on Maintainable Software Practices in
  e-Science}, Chicago, IL, USA, Oct. 2012.
\newblock Position paper.

\bibitem{ref:darlington95}
J.~Darlington, Y.~Guo, H.~W. To, and J.~Yang.
\newblock Functional skeletons for parallel coordination.
\newblock In {\em EURO-PAR'95 Parallel Processing}, pages 55--69.
  Springer-Verlag, 1995.

\bibitem{ref:milepostgcc}
{G. Fursin et al.}
\newblock Milepost gcc: Machine learning enabled self-tuning compiler.
\newblock {\em International Journal of Parallel Programming}, 39(3):296--327,
  2011.

\bibitem{ref:ukescience}
T.~Hey and A.~E. Trefethen.
\newblock The uk e-science core programme and the grid.
\newblock In P.~M.~A. Sloot, A.~G. Hoekstra, C.~J.~K. Tan, and J.~J. Dongarra,
  editors, {\em Computational Science -- ICCS 2002}, volume 2329 of {\em LNCS},
  pages 3--21. Springer Berlin Heidelberg, 2002.

\bibitem{ref:WoHaFe13}
{K. Wolstencroft et al.}
\newblock The {Taverna} workflow suite: designing and executing workflows of
  {Web Services} on the desktop, web or in the cloud.
\newblock {\em Nucleic Acids Research}, 2013.
\newblock First published online May 2, 2013, doi:10.1093/nar/gkt328.

\bibitem{ref:opencl}
{Khronos Group}.
\newblock {OpenCL\texttrademark -- The open standard for parallel programming
  of heterogeneous systems}.
\newblock http://www.khronos.org/opencl/, 2013.
\newblock accessed 1$^{st}$ September 2013.

\bibitem{ref:op2}
G.~R. Mudalige, M.~B. Giles, I.~Reguly, C.~Bertolli, and P.~H.~J. Kelly.
\newblock {OP2: A}n active library framework for solving unstructured
  mesh-based applications on multi-core and many-core architectures.
\newblock In {\em Innovative Parallel Computing (InPar), 2012}, pages 1--12,
  2012.

\bibitem{ref:rackspace}
{Rackspace Limited}.
\newblock {Cloud Overview | Rackspace Hosting}.
\newblock http://www.rackspace.co.uk/cloud/, 2013.
\newblock accessed 1$^{st}$ September 2013.

\bibitem{ref:oski}
R.~Vuduc, J.~W. Demmel, and K.~A. Yelick.
\newblock {OSKI}: A library of automatically tuned sparse matrix kernels.
\newblock {\em Journal of Physics: Conference Series}, 16(1):521--530, 2005.

\end{thebibliography}
\end{document}